\documentstyle[11pt,paspconf,epsfig]{article}

\begin{document}

\title{Radiation from Relativistic Strongly Magnetized Outflows}

\author{Vladimir V. Usov}
\affil{Department of Condensed Matter Physics, Weizmann Institute,
Rehovot 76100, Israel}



\begin{abstract}
Relativistic strongly magnetized winds outflowing from 
fast-rotating compact objects like millisecond pulsars
with surface magnetic fields of $\sim 10^{15}-10^{16}$ 
G are plausible sources of cosmological $\gamma$-ray bursts.
In such winds, there are at least three regions where extremely 
powerful X-ray and $\gamma$-ray  emission may be  generated.
The first radiating region is the wind photosphere that is
at a distance of $\sim 10^9$ cm from the compact object. The second 
radiating region is at a distance of $\sim 10^{13}-10^{14}$ cm.
In this region, the striped component of the wind field is transformed into 
large-amplitude electromagnetic waves. The third
radiating region is at a distance of $\sim 10^{16}-10^{17}$ cm,
where deceleration of the wind due to its interaction with an ambient 
medium becomes important. Radiation from all these regions is considered.
\end{abstract}




\section{Introduction}
\label{Introduction} 

Many ideas about the nature of $\gamma$-ray bursts have
been discussed during last 25 years after the burst discovery (for
a review, see Blaes 1994, Harding 1994, Hartman 1995; 
Dermer \& Weiler 1995, Fishman \& Meegan 1995, Greiner 1998, Piran 1998).
Among these ideas, there was a suggestion that 
the sources of $\gamma$-ray bursts (GRBs) are at cosmological 
distances, i.e. at a redshift $z\sim 1$ (Usov \& Chibisov 1975,
van den Berg 1983, Paczy\'nski 1986, Goodman 1986,
Eichler et al. 1989). After the BATSE data became available 
(Meegan et al. 1992, 1994), 
the idea of a cosmological origin of GRB sources has come 
to be taken very seriously (e.g., Paczy\'nski 1991, Fishman \& Meegan 1995).
Recent detections of absorption and emission features at a redshift
$z=0.835$ in the optical afterglow of GRB 970508 
(Metzger et al. 1997) and at redshift $z=3.42$ in the host galaxy of GRB 
971214 (Kulkarni et al. 1998) clearly demonstrate that at least
some of the GRB sources lie at cosmological distances.
A common feature of all
acceptable models of cosmological $\gamma$-ray bursters is a
relativistic wind as the source of GRB radiation. The Lorentz factor,
$\Gamma_0$, of such a wind is about $10^2-10^3$ or
even more (e.g., Fenimore, Epstein \& Ho 1993, Baring \& Harding 1997).
A very strong magnetic field may be in the plasma 
outflowing from cosmological $\gamma$-ray bursters (Usov 1992, 
1994a,b, Thompson \& Duncan 1993, Blackman, Yi \& Field 1996, Vietri 1996,
Katz 1997, M\'esz\'aros \& Rees 1997a, Dai \& Lu 1998). 
Below, we discuss both thermal and non-thermal radiation from
relativistic strongly magnetized winds that are plausible sources of 
cosmological $\gamma$-ray bursters.

\section{Relativistic strongly magnetized winds from cosmological $\gamma$-ray 
bursters and their radiation: a plausible scenario}
\label{model} 

The energy output of cosmological $\gamma$-ray bursters in 
$\gamma$-rays typically is $10^{51}-10^{53}$ ergs (e.g., Wickramasinghe et al. 
1993, Tamblyn \& Melia 1993, Lipunov et al. 1995) and may be as high as 
$3\times 10^{53}$ ergs (Kulkarni et al. 1998) or even more (Kulkarni et al. 
1999). These estimates assume isotropic emission of GRBs.
Such a high energetics of cosmological $\gamma$-ray bursters
and a short time scale of $\gamma$-ray flux variability
call for very compact objects as sources of GRBs (Hartmann 1995,
Piran 1998 and references therein). These objects may
be either millisecond pulsars which are arisen from accretion-induced
collapse of white dwarfs in close binaries (Usov 1992) or differentially
rotating disk-like objects which are formed by the merger of a binary
consisting of two neutron stars (Eichler et al. 1989, Narayan, Paczy\'nski
\& Piran 1992).
Such very young fast-rotating compact objects have two possible sources
of energy which may be responsible for the radiation of cosmological GRBs. 
These are the thermal energy of the compact objects and the kinetic energy
of their rotation. The thermal energy may be transformed into
$\gamma$-rays by means of the following sequence of processes (for review, 
see Piran 1998): (1) Emission of neutrinos and cooling of the object;
(2) Absorption of neutrinos $(\nu_i +
\bar\nu_i \rightarrow e^+ + e^- )$ and formation of a fireball which mainly
consists of electrons and positrons;
(3) Expansion of the fireball and formation of a relativistic shell, $\Gamma_0
> 10^2$; (4) Interaction of the shell with an external medium and
acceleration of electrons to very high energies; and (5) 
Generation of $\gamma$-rays by highly accelerated electrons. 

The maximum thermal
energy, $Q_{\rm th} ^{\rm max}$, of very young compact objects  
is high enough to explain the energy output
of cosmological GBBs, $Q_{\rm th} ^{\rm max}\simeq$ a few $\times 10^{53}$ ergs.
However, the fraction of the thermal energy that
is converted into the energy of the electron-positron
fireball and then into the kinetic energy of the relativistic shell is very small
and cannot be essentially more than $10^{-3}-10^{-2}$ (Goodman, Dar \&
Nussinov 1987, Eichler et al. 1989, Janka \& Ruffert 1996,
Piran 1998). Moreover, the efficiency of
transformation of the kinetic energy of a relativistic shell into radiation 
cannot be more than $30-40$\% (Blandford \& Eichler 1987). 
Therefore, neutrino powered winds outflowing from very young compact objects
may be responsible for the radiation of cosmological GRBs only if they are 
well collimated, with opening angle about a few degrees or even less.
For both neutron stars and
post-merger objects, such a collimation of neutrino powered winds
is very questionable (Woosley 1993, Piran 1998  and references therein). 

The rotational energy of compact objects at the moment of their formation
may be comparable with the thermal energy, $Q_{\rm rot}
^{\rm max}\simeq  Q_{\rm th} ^{\rm max}$. The
efficiency of transformation of the rotational energy to
the energy of a relativistic strongly magnetized wind and then
to the energy of high-frequency radiation may be as high as
almost 100\% (see Usov 1994a,b; Blackman et al. 1996 and below).
For some time the theoretical expectation has been that
rotation powered neutron stars (pulsars)
should generate collimated outflows (e.g., Benford 1984,
Michel 1985, Sulkanen \& Lovelace 1990). The Crab, Vela, PSR B1509-58
and possible PSR B1951+32 all show evidence that this is indeed
the case (Hester 1998, Gaensler et al. 1999 and references therein).
If the energy flux from the source of GRB 990123
in the direction to the Earth is only about ten
times more than the energy flux averaged over all directions,
the model of GRBs based on the rotation powered winds
can easily explain the energetics of
such an extremal event as GRB 990123 (Kulkarni et al. 1999). Such an 
anisotropy of emission from the burst sources  doesn't contradict 
available data on GRBs (e.g., Perna \& Loeb 1998). 
In the case of typical GRBs with the energy output of
$10^{51}-10^{53}$ ergs, this model can explain their energetics even if
the emission of GRBs is nearly isotropic.
Therefore, the rotational energy of compact
objects is a plausible source of energy for cosmological GRBs,
not the thermal energy.
 
In many papers (e.g., Usov 1992, Thompson \& Duncan 1993,
Blackman et al. 1996; Klu\'zniak \& Ruderman 1998), it was argued that
the strength of the magnetic field $B_{_S}$ at the surface of compact objects 
may be as high as $\sim 10^{16}$~G or even more.  Such a strong magnetic field 
leads to both deceleration of the rotation of the compact object
on a time  scale of seconds and generation of a strongly
magnetized wind that flows away from the object at 
relativistic speeds, $\Gamma_0\simeq 10^2-10^3$ (e.g., Usov 1994a).
The outflowing wind is Poynting flux$-$dominated, i.e., $\sigma = L_{\pm}/
L_{_P}\ll 1$, where 

\begin{equation}
L_{_P}\simeq {2\over 3} {B_{_S}^2R^6\Omega^4\over c^3}\simeq 2\times
10^{52}\left({B_{_S}\over  10^{16}\,\,{\rm G}}
\right)^2\left({R\over 10^6\,\,{\rm cm}}\right)^6
\left({\Omega\over 10^4\,\,{\rm s}^{-1}}\right)^4\,\,
{\rm ergs\,\,s}^{-1}
\end{equation}

\noindent
is the luminosity of the compact object in the Poynting flux, $ L_{\pm}$ is its 
luminosity in both electron-positron pairs and radiation, $c$ is the speed of 
light, $R$ is the radius of the compact object
and $\Omega$ is its angular velocity; $R\sim 10^6$ cm and $\Omega\sim 
10^4$ s$^{-1}$ for both millisecond pulsars and post-merger objects. 
For compact objects with extremely strong magnetic fields, 
$B_{_S}\sim 10^{16}$ G, it is expected that $\sigma $ is $\sim 0.01-0.1$ 
(Usov 1994a).

\begin{figure}
\centerline{\epsfig{file=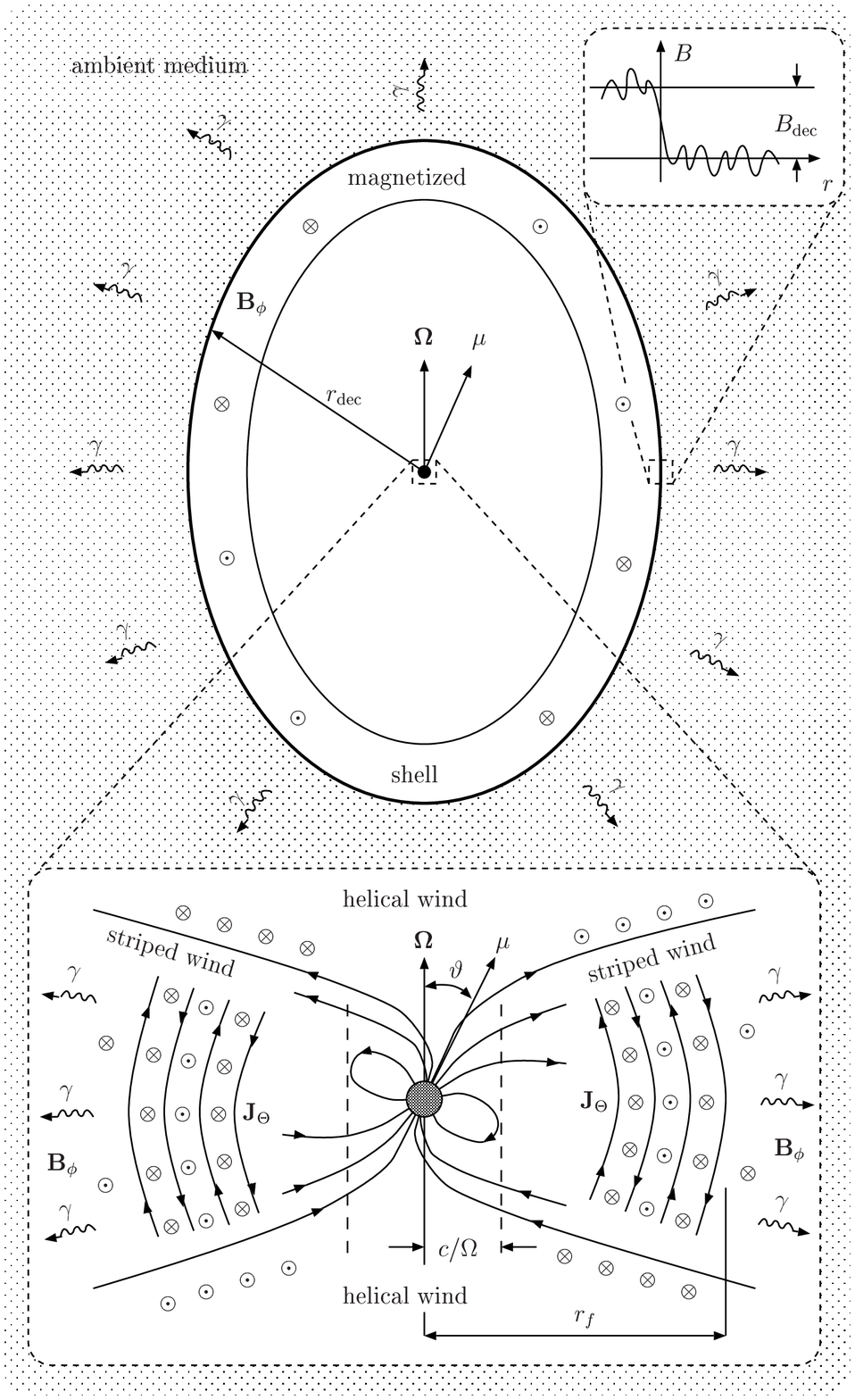,width=14cm,height=14cm}}
\caption{Sketch (not to scale) of the region where a cosmological
GRB is generated by a relativistic, strongly magnetized wind. This wind
is produced by a fast-rotating compact object like a millisecond pulsar with
the angular velocity ${\bf \Omega}$ and the magnetic moment ${\bf \mu}$.} 
\label{Fig1_u}
\end{figure}

A plausible magnetic topology for a relativistic magnetized 
wind outflowing from an oblique rotator ($\vartheta\neq 0$) with a 
nearly dipole magnetic field is shown in Figure~\ref{Fig1_u}, 
where $\vartheta$ is the angle between the rotational 
and magnetic axes. Near the rotational poles, 
the wind field should be helical (e.g., Coroniti 1990). 
This is because the magnetic flux originates in a single polar cap.
Near the rotational equator, the toroidal magnetic 
field of the wind should be striped and alternates in polarity 
on a scalelength of $\pi (c/\Omega )\sim 10^7$ cm. 
These magnetic stripes are separated by thin current sheets $(J_\theta)$. 
Off the equator, the magnetic flux in the toward and away stripes is 
unequal if $\vartheta\neq  \pi/ 2$. In other words,
in the striped region, the wind field is a superposition of a pure
helical field and a pure striped field with nearly equal magnetic
fluxes in adjacent stripes.

Since the luminosity of a $\gamma$-ray
burster in a relativistic magnetized wind drops in time, $L_{_P} 
\propto t^{-\beta}$, the wind structure at the moment  $t\gg\tau_{_\Omega}$ 
is similar to a shell with the radius $r\simeq ct$, where $\beta$ is a 
numerical index, $1\leq \beta\leq 2$, and
$\tau_{_\Omega}\sim 10^{-2}-10^2$ s is the characteristic time 
of deceleration of the compact object rotation because of the action of 
the electromagnetic torque and the torque connected with 
the generation of gravitational radiation (Usov 1992, Yi \& Blackman 1998).
The thickness of the shell is $\sim c\tau_{_\Omega}$. 
The strength of the magnetic field in the shell is about

\begin{equation}
B\simeq B_{_S}\frac {R^3}{r_{\rm lc}^2r}
\simeq  10^{15}\frac Rr\left(
\frac{B_{_S}}{ 10^{16}\,\rm{G}}\right) \left(\frac\Omega
{10^4\,\rm{s}^{-1}}\right)^2\rm{G},
\label{Bobj}
\end{equation}
 
\noindent
where $r_{\rm lc}=c/\Omega=3\times 10^6(\Omega /10^4\,$s$^{-1})$ cm
is the radius of the light cylinder. 

For relativistic, strongly magnetized winds with
$L_{_P}\sim 10^{52}$ ergs s$^{-1}$ and $\sigma \sim 0.01-
0.1$, the optical depth of the outflowing electron-positron plasma
near the compact objects is very high, up to $\sim 10^{12}-10^{13}$
(Usov 1994a). In this case, the electron-positron plasma and 
radiation are in quasi-thermodynamic equilibrium.
During outflow, the electron-positron plasma 
accelerates and its density decreases (e.g., Paczy\'nski 1986,
Goodman 1986). At a distance $r_{\rm ph}$ from the compact object,
where the optical depth for the bulk of the photons is $\sim 1$,
the radiation propagates freely. The radius of the
wind photosphere is $r_{\rm ph}\sim 10^9$
cm, and the Lorentz factor of the outflowing plasma at 
the photosphere is $\sim 10^2$ (see \S~3).

At the distance

\begin{equation}
r < r_f\simeq 2\times 10^{14} \sigma^{3/4}\left(
{B_{_ S}\over  10^{16}\,{\rm G}}\right)^{1/2}
\left({\Omega\over 10^4\,{\rm s}^{-1}}\right)^{1/2}\,\,
{\rm cm}\,,
\label{rnth}
\end{equation}

\noindent
the magnetic field of the wind is frozen in the outflowing 
plasma (Usov 1994a, Blackman \& Yi 1998), and there is no 
reason for powerful non-thermal radiation to be generated.

At $r > r_f$, the wind density is not sufficient 
to screen displacement currents, and the striped component of
the wind field is transformed into large-amplitude
electromagnetic waves (LAEMWs) (Usov 1975, 1994a,b, Blackman 
et al. 1996, Melatos \& Melrose 1996a,b).  
Outflowing particles are accelerated in the field of LAEMWs 
and generate powerful synchro-Compton radiation (see \S~4).

At $r\gg r_f$, the magnetic field is helical 
everywhere in the  outflowing wind (see Fig. 1). Such a relativistic strongly 
magnetized wind expands more or less freely up to the distance 

\begin{equation}
r_{\rm{dec}}\simeq 5\times 10^{16}\left({\frac{Q_{\rm{kin}}}{10^{52}\,
\rm{ergs}}}\right)^{1/3}\left({\frac n{1\,\rm{cm}^{-3}}}\right)^{-1/3}\left(
{\frac{\Gamma_0}{10^2}}\right)^{-2/3}\rm{cm},
\label{rdec}
\end{equation}

\noindent at which 
deceleration of the wind due to its interaction with an ambient
medium becomes important (Rees \& M\'{e}sz\'{a}ros 1992), where
$n$ is the density of the ambient medium and $Q_{\rm{kin}}$
is the kinetic energy of the outflowing wind, $Q_{\rm{kin}}
\leq Q_{\rm rot}\leq Q^{\rm max}_{\rm rot}$ .  

It was suggested by M\'{e}sz\'{a}ros and Rees (1992, 1993) that in the process 
of the wind $-$ ambient medium interaction at $r\sim r_{\rm{dec}}$, 
an essential part of the wind energy may be transferred to 
high-energy electrons and then to high-frequency (X-ray and
$\gamma$-ray) emission. This suggestion was recently confirmed by
numerical simulations (see Smolsky \& Usov 1996, 1999, Usov \& Smolsky
1998 and \S~5). 

Summarizing, in relativistic strongly magnetized winds there are three 
regions where powerful radiation of GRBs may be generated. 
The first radiating region is the wind photosphere that is
at the distance $r_{\rm ph}\sim 10^9$ cm from the compact object. The second 
radiating region is at the distance $r_f \sim 10^{13}-10^{14}$ cm.
In this region, the striped component of the wind field is transformed into 
LAEMWs. The third radiating region is at the distance $r_{\rm dec}
\sim 10^{16}- 10^{17}$ cm where deceleration of the wind due to its 
interaction with the ambient medium becomes important. 
Below, we consider radiation from all these regions.

\section{Thermal radiation from the wind photosphere}

For a spherical optically thick electron-positron wind,
the radius of the photosphere is (Paczy\'nski 1990)

\begin{equation}
r_{\rm ph}\simeq \left({L_\pm\over 4\pi aT_{_0}^4
\Gamma_{\rm ph}^2}\right)^{1/2}\,,
\end{equation}

\noindent
where $T_{_0}$ is the temperature of the electron-positron plasma 
at $r\simeq r_{\rm ph}$ in the comoving frame, $a=5.67\times 10^{-5}$
ergs cm$^{-2}$ K $^{-4}$ s$^{-1}$ is the Stefan-Boltzmann constant, and
and $\Gamma_{\rm ph}$ is the mean Lorentz factor of the ourflowing 
plasma at $r\simeq r_{\rm ph}$.

Since $\Gamma_{\rm ph}\simeq r_{\rm ph}/r_{\rm lc}$ and $T_{_0}\simeq
2\times 10^8$ K (e.g., Paczy\'nski 1986, Goodman 1986), from equation (5)
we have (Usov 1994a)

\begin{equation}
r_{\rm ph}\simeq \left({L_\pm r_{\rm lc}^2 \over 4\pi aT_{_0}^4}\right)^{1/4}
\simeq 6\times 10^8\sigma^{1/4}\left({B_{_S}\over  10^{16}\,\,{\rm G}}
\right)^{1/2}\left({\Omega\over 10^4\,\,{\rm s}^{-1}}\right)^{1/2}\,\,{\rm cm}\,,
\end{equation}

\begin{equation}
\Gamma_{\rm ph}\simeq 2\times 10^2\sigma^{1/4}\left({B_{_S}\over 10^{16}\,
\,{\rm G}}\right)^{1/2}\left({\Omega\over 10^4\,\,{\rm s}^{-1}}\right)^{3/2}\,.
\end{equation}

At $r < r_{\rm ph}$ the optical depth increases sharply with decreasing $r$.
Therefore, any energy that is inherited by particles and radiation at the
distance to the compact object a few times smaller than $r_{\rm ph}$
will be thermalized before it is radiated at $r\simeq r_{\rm ph}$.

The temperature that corresponds to the blackbody-like
radiation from the wind photosphere is

\begin{equation}
T_{\rm th}\simeq 2\Gamma_{\rm ph}T_{_0}
\simeq 8\times 10^{10}\sigma^{1/4}\left({B_{_S}\over  10^{16}\,\,{\rm G}}
\right)^{1/2}\left({\Omega\over 10^4\,\,{\rm s}^{-1}}\right)^{3/2}\,\,{\rm K}\,.
\end{equation}

\noindent
The mean energy of thermal photons is $\langle\varepsilon_{\rm th}\rangle 
\simeq 3kT_{\rm th}$. For typical parameters
of the wind sources, $B_{_S}\simeq 10^{16}$ G, $\Omega\simeq 10^4$ 
s$^{-1}$ and $\sigma\simeq 0.01-0.1$, we have
$\langle\varepsilon_{\rm th}\rangle \simeq 1$ MeV within a factor of 2-3 or so.

For the outflowing electron-positron wind, the flux of particles at 
$r > r_{\rm ph}$ is (Paczy\'nski 1990)

\begin{equation}
\dot N_\pm\simeq {4\pi c r_{\rm ph}\Gamma_{\rm ph}^2\over \sigma_{_T}}\,,
\end{equation}

\noindent
where $\sigma_{_T}=6.65\times 10^{-25}$ cm$^2$ is the Thomson 
cross-section. Equations (6), (7) and (9) yield

\begin{equation}
\dot N_\pm\simeq
 10^{49}\sigma^{3/4}\left({B_{_S}\over 10^{16}\,\,{\rm G}}
\right)^{3/2}\left({\Omega\over 10^4\,\,{\rm s}^{-1}}\right)^{7/2}\,\,{\rm s}^{-1}\,.
\end{equation}

\noindent
From equations (1), (7) and (10), we can see that
the total energy flux of the wind in electron-positron pairs at $r>r_{\rm ph}$
is a very small part of the luminosity of the compact object in both 
electron-positron pairs and radiation at the moment of its formation,
$m_ec^2\Gamma_{\rm ph}\dot N_\pm\simeq 10^{-7}L_\pm$,
where $m_e$ is the electron mass. Therefore,
the thermal luminosity, $L_{\rm th}$, of the wind photosphere practically 
coincides with $L_\pm$. 

\section{Non-thermal radiation from the region of transformation of 
the striped wind component into LAEMWs}

At $r\simeq r_{\rm ph}$, the density of the outflowing electron-positron 
plasma,

\begin{equation}
n_\pm = {\dot N_\pm\over 4\pi c r_{\rm ph}^2}\simeq
7\times 10^{19}\sigma^{1/4}
\left({B_{_S}\over 10^{16}\,\,{\rm G}}
\right)^{1/2}\left({\Omega\over 10^4\,\,{\rm s}^{-1}}\right)^{5/2}
\,\,{\rm cm}^{-3}\,,
\end{equation}

\noindent
is essentially higher than the critical value (e.g., Goldreich \& 
Julian 1969)

\begin{equation}
n_{\rm cr}={\Omega B\over 4\pi c e}\simeq 2\times 10^{14}\sigma^{-1/4}
\left({B_{_S}\over 10^{16}\,\,{\rm G}}
\right)^{1/2}\left({\Omega\over 10^4\,\,{\rm s}^{-1}}\right)^{5/2}
\,\,{\rm cm}^{-3}\,.
\end{equation}

\noindent
In this case, the magnetic field is frozed in the wind plasma.
In the process of the plasma outflow, the plasma density
decreases in proportion to $r^{-2}$ while
the critical density is $\propto r^{-1}$. At the distance
$r_f$ that is given by equation (3), the plasma density
is equal to the critical density, $n_\pm =n_{\rm cr}$.
At $r > r_f\sim 10^{13}-10^{14}$~cm, the plasma density is not sufficient 
to screen displacement currents, and the striped component of
the wind field is transformed into LAEMWs due to development of 
magneto-parametric instability (Usov 1975, 1994a,b, Blackman 
et al. 1996, Melatos \& Melrose 1996a,b). The frequency 
of generated LAEMWs is equal to $\Omega$, and their amplitude is 
$\sim B$. Outflowing particles are accelerated in the field of LAEMWs 
to the Lorentz factor of $\sim (L_{_P}/m_ec^2\dot N_\pm)^{2/3}\sim
10^6$ and generate non-thermal
synchro--Compton radiation with the typical energy of photons
$\langle\varepsilon_\gamma\rangle\sim \hbar\Omega(L_{_P}/
m_ec^2\dot N_\pm)^2\sim 1$ MeV (Usov 1994a,b,
Blackman et al. 1996, Blackman \& Yi 1998), where $\hbar\simeq
10^{-27}$ ergs s is the Planck constant. 
[For a review on acceleration of electrons in the fields of LAEMWs 
and synchro-Compton radiation
see (Gunn \& Ostriker 1971, Blumenthal \& Tucker 1974).]

The radiative damping length for LAEMWs generated at $r\sim r_f$
is a few orders of magnitude less than $r_f$ (Usov 1994a). Therefore,
at $r\gg r_f$ LAEMWs decay 
almost completely, and their energy is transferred to 
high-energy electron-positron pairs and then to X-ray and $\gamma$-ray 
photons. It is worth noting that in the case when the magnetic 
axis is perpendicular to the rotational axis,  $\vartheta = \pi /2$, the 
electromagnetic field of the Poynting flux--dominated wind is purely
striped just as vacuum magnetic dipole waves (Michel 1971),
and almost all energy of the wind is radiated in X-rays and
$\gamma$-rays at $r\sim r_f$  (Usov 1994a, Blackman et al. 1996, 
Blackman \& Yi 1998). In this case, the total energy output in 
hard photons per a GRB may be as high as $Q^{\rm max}_{\rm rot}
\simeq$ a few $\times 10^{53}$ ergs.

At $\vartheta\simeq \pi /2$, when the bulk of the 
wind energy is transferred into $\gamma$-rays at $r\sim r_f$ and
the residual energy of the wind at $r\gg r_f$ is small,
afterglows which are generated at $r > r_{\rm dec}\gg r_f$ and
accompany GRBs are weak irrespective of that the GRBs themselves 
may be quite strong. This may explain the fact that X-ray, optical and
radio afterglows have been observed in some strong GRBs but not 
in others (e.g., Piran 1998 and references therein).

\section{Non-thermal radiation from the region of the wind --
ambient medium interaction}

For consideration of the interaction between a
relativistic magnetized wind and an ambient medium, it is convenient
to switch to the co-moving frame of the outflowing plasma (the wind frame).
While changing the frame, the magnetic and electric fields in the wind
are reduced from $B$ and $E=B[1-(1/\Gamma_0^2)]^{1/2}
\simeq B$ in the frame of
the $\gamma$-ray burster to $B_0\simeq B/\Gamma_0$ and $E_0=0$
in the  wind frame. Using this and equations (2) and (4), for 
typical parameters of both cosmological $\gamma$-ray bursters, $B_{_S}
\simeq 10^{16}$ G, $\Omega \simeq 10^4$ s$^{-1}$, $Q_{\rm kin}\simeq
10^{52}-10^{53}$ ergs and $\Gamma_{_0}\simeq 10^2-10^3$, and 
the ambient medium, $n\simeq 1-10^2$ cm$^{-3}$, 
we have that at $r\simeq r_{\rm dec}$ the strength of the magnetic field 
at the wind front is $B_{\rm dec} \simeq 10^4-4\times 10^5$ G in 
the burster frame and $B_0\simeq B_{\rm dec}/\Gamma_{_0}
\simeq 10^2-10^3$ G in the wind frame.

In the wind frame, the problem of the
wind $-$ ambient medium interaction is identical to the problem of
collision between a wide relativistic beam of cold plasma and a region
with a strong magnetic field which is called a magnetic barrier.
Recently, the interaction of a wide relativistic plasma beam with a magnetic 
barrier was studied numerically (Smolsky \& Usov 1996, 1999, Usov
\& Smolsky 1998).
In these studies, the following initial condition of the beam $-$ barrier 
system was assumed.  Initially, at $t=0$,
the ultrarelativistic homogenious neutral beam of protons and electrons 
(number densities $n_p=n_e\equiv n_0$) runs along the $x$ axis and 
impacts at the barrier, where $n_0$ is constant. The beam is infinite 
in the $y-z$ dimensions and semi-infinite in the $x$ dimension.
The  magnetic field of
the barrier ${\bf B}_0$ is uniform and transverse to the beam velocity,
${\bf B}_0=B_0{\bf \hat e}_z\Theta[x]$, where 
$B_0$ is constant and $\Theta [x]$ is the step function equal to
unity for $x>0$ and to zero for $x<0$.  At the front of the barrier, 
$x=0$, the surface current $J_y$ runs along the $y$ axis
to generate the jump of the magnetic field. The value of this 
current per unit length of the front across the current direction is 
$cB_0/4\pi$. A 1${1\over 2}$D time-dependent solution for the problem
of the beam $-$ barrier interaction was constructed, i.e.,
electromagnetic fields (${\bf E}=E_x{\bf \hat
  e}_x+E_y{\bf \hat e}_y$; ${\bf B}=B{\bf \hat e}_z$) and motion of 
the beam particles in the $x-y$ plane were found. The field 
structure and the beam particle motion were treated self-consistently
except the external current $J_y$ which is fixed in 
our simulations. 

The main results of our simulations are the following
(Smolsky \& Usov 1996, Usov \& Smolsky 1998).

1. When the energy densities of the beam and the magnetic field,
${\bf B}_0$, of the barrier are comparable, 

\begin{equation}
\alpha = 8\pi n_0m_pc^2(\Gamma_0-1)/B^2_0\sim 1\,, 
\label{alpha}
\end{equation}

\noindent 
where $m_p$ is the proton mass, the process
of the beam $-$ barrier interaction is strongly nonstationary, and the 
density of outflowing protons after their reflection from the barrier
is strongly non-uniform. The ratio of the maximum density of
reflected protons and their minimum density is $\sim 10$.
 
\begin{figure}
\centerline{\epsfig{file=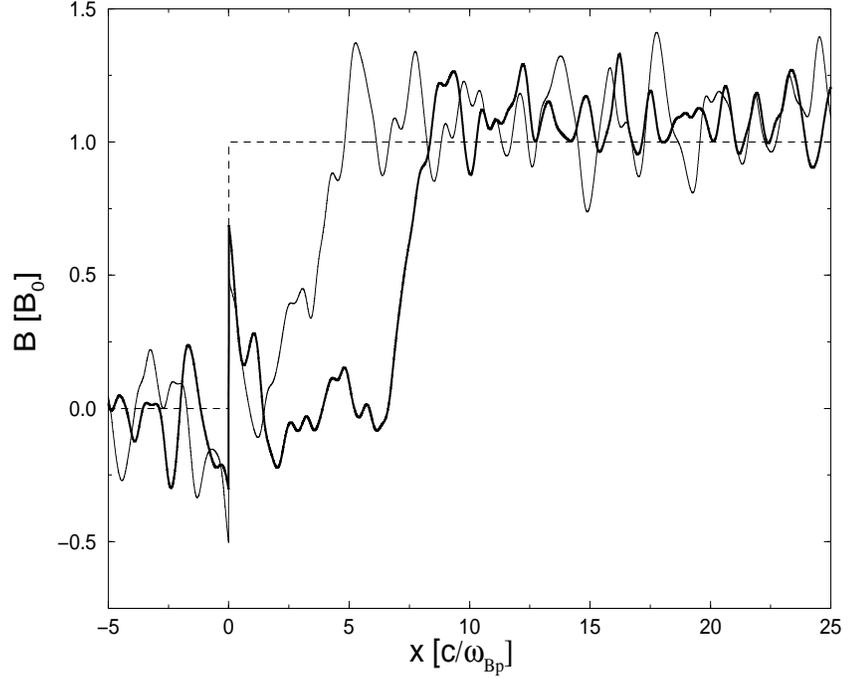,angle=-90,width=11cm,height=9cm}}
\caption{Distribution of magnetic field for a simulation
with $B_0=300$ G, $\Gamma_0=300$ and $\alpha =2/3$ at
the moments $t=0$ (dotted line), $t=7.96T_p$ (thin solid line),
and $t=15.9T_p$ (thick solid line). $T_p=2\pi / \omega_{Bp}$ is 
the proton gyroperiod in the magnetic field, $B_0$, of the barrier.
} \label{fig2_usov}
\end{figure}

2. At $\alpha > \alpha _{\rm cr}\simeq 0.4$, 
the depth of the beam particle penetration into the barrier 
increases in time, $x_{\rm pen}\simeq v_{\rm pen} t$, where 
$v_{\rm pen}$ is the mean velocity of the penetration
into the barrier. The value of $v_{\rm pen}$ is subrelativistic and 
varies from zero (no penetration)  at $\alpha \leq \alpha_{\rm cr}$ to
$0.17c$ at $\alpha =1$ and $0.32 c$ at $\alpha =2$.
At $\alpha > \alpha _{\rm cr}$,
the magnetic field of the barrier at the moment $t$ roughly
is $B(t)\simeq B_0\Theta[x-x_{\rm {pen}}(t)]$ (see Fig.~\ref{fig2_usov}).
In other words,  the front of the beam -- barrier interaction
is displaced into the barrier with the velocity $v_{\rm pen}$.
For $\alpha > \alpha _{\rm cr}$, our consideration of the beam -- barrier 
interaction in the vicinity of the new front, $x\simeq x_{\rm {pen}}$, 
is {\it completely self-consistent}, and no 
simplifying assumptions besides geometrical ones are exploited.

3. At the front of the barrier, $x\simeq x_{\rm {pen}}(t)$, the surface 
current varies in time because of strong nonstationarity 
of the beam $-$ barrier interaction at $\alpha\sim 1$,
and LAEMWs are generated (Fig. 2).
The frequency of these waves is about the proton 
gyrofrequency $\omega_{Bp}=eB_0/m_pc\Gamma_0$
in the field of the barrier $B_0$. The wave amplitude 
$ B_w$ can reach $\sim 0.2 B_0$.

\begin{figure}
\centerline{\epsfig{file=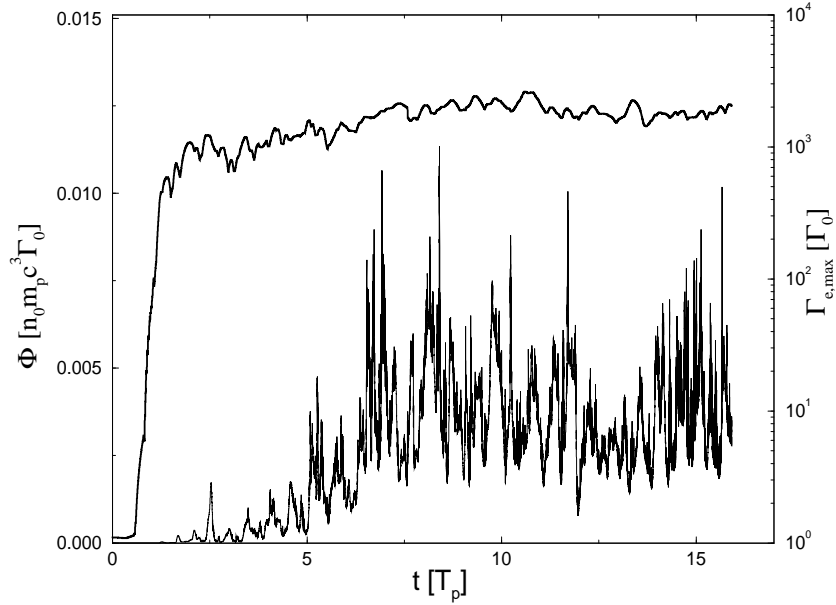,angle=-90,width=11cm,height=8cm}}
\caption{Maximum energy of accelerated electrons (thick line)
and intensity of their synchrotron radiation per unitary 
area of the front of the magnetic barrier (thin line) in a simulation
with $B_0=300$ G, $\Gamma_0=300$ and $\alpha =2/3$.
} \label{fig3_usov}
\end{figure}

4. At $\alpha \sim 1$, strong electric fields are generated in the vicinity 
of the front of the barrier, $x\simeq x_{\rm {pen}}(t)$, and electrons of 
the beam are accelerated in these fields up to the mean energy 
of protons, i.e. up to $\sim m_pc^2 \Gamma_0$ (Fig.~\ref{fig3_usov}). 
At $\alpha_{\rm cr} < \alpha < 1$,
the mean Lorentz factor of outflowing high-energy electrons 
after their reflection and acceleration at the barrier front
depends on $\alpha$ and is 

\begin{equation}
\langle\Gamma_e^{\rm {out}}\rangle \simeq 0.2
\left(\frac{m_p}{m_e}\right)\Gamma_0
\label{Gammamean} 
\end{equation}

\noindent 
within a factor of 2. The total energy of the accelerated electrons is
about 20\% of the energy in the outflowing protons which are
reflected from the magnetic barrier. 

5. At $\alpha_{\rm cr} < \alpha < 1$, the mean Lorentz 
factor of protons reflected from the barrier is
$\langle \Gamma_p^{\rm{out}}\rangle \simeq (0.7\pm 0.1)\Gamma_0$,
i.e. the process of the beam proton reflection from the barrier
is non-elastic, and about  30\% of the initial kinetic energy 
of the beam protons is lost in
this collision. The energy that is lost by the beam protons 
is transferred to high-energy electrons and 
LAEMWs. 
Typically, the energy in these waves is a few
times smaller than the energy in high-energy electrons. 

In the burster frame, a magnetized wind flows away
from the burster at relativistic speeds and collides with an ambient
medium. In the process of such a collision, the outflowing wind loses its 
energy. From the listed results of our simulations 
of the beam -- barrier collision (Smolsky \& Usov 1996, 
Usov \& Smolsky 1998), it follows that at $r\sim r_{\rm dec}$,
where $\alpha$ is $\sim 1$, about 70\% of the energy losses of
the wind are transferred to protons of the ambient medium
which are reflected from the wind front. The mean
energy of reflected protons is about $m_pc^2\Gamma_0^2$. The other
30\% of the wind energy losses are distributed
between high-energy electrons and LAEMWs. 
As a rule, the total energy in accelerated electrons
is a few times higher than the total energy in LAEMWs.

High-energy electrons accelerated at the wind front generate
non-thermal radiation while they move in both the magnetic field of
the wind and the electromagnetic fields of LAEMWs
that propagate ahead of the wind front.  Let us consider this radiation.

\subsection{Synchrotron radiation from the wind front}

In our simulations of the beam $-$ magnetic barrier interaction (Smolsky \& 
Usov 1996, Usov \& Smolsky 1998), the examined space-time domain is

\begin{equation}
x_{\rm min}<x<x_{\rm max},\,\,\,\,\,\,\,\,\,\,0<t<t_{\rm max},
\label{xt}
\end{equation}

\noindent
where $x_{\rm max}$ and $-x_{\rm min}$ are equal to a few $\times (1-10)
(c/\omega_{Bp})$, $t_{\rm max}$ is equal to a few $\times (1-10)T_p$, and
$(c/\omega_{Bp})$ is the proton gyroradius in the magnetic field of 
the barrier $B_0$. Non-thermal radiation of high-energy 
electrons from the examined space domain was calculated for the beam 
and barrier parameters which are relevant to cosmological GRBs (Smolsky 
\& Usov 1996, Usov \& Smolsky 1998).  Figure 3 shows the intensity of 
radiation as a function of time $t$ for a simulation with $B_0=300$ G,
$\Gamma_0=300$, $\alpha =2/3$, $x_{\rm min}=-5(c/\omega_{Bp})$
and $x_{\rm max}=30(c/\omega_{Bp})$. In all our simulations, the bulk of 
calculated radiation is generated via synchrotron mechanism
in a compact vicinity, $x_{\rm pen}- 2 (c/\omega_{Bp}) <x
< x_{\rm pen}$, of the barrier front where both the strength of the magnetic
field is of the order of $B_0$ and the mean energy of accelerated 
electrons is extremely high. Radiation of high-energy
electrons in the fields of LAEMWs is negligible
($\sim 1$\% or less) because both the fields of these waves are about
an order of magnitude smaller than $B_0$ (see below) and
the length of the examined space domain (15)
is restricted for computational reasons.  

At $\alpha\sim 1$, the mean energy of
synchrotron photons generated at the front of the barrier is

\begin{equation}
\langle \varepsilon _\gamma \rangle \simeq 0.1\left( {\frac{\Gamma
_0}{10^2}} \right) ^2\left( {\frac{B_0}{10^3\,\rm{G}}}\right) \;\;
\rm{MeV}\,. 
\label{eps}
\end{equation}
 
\noindent 
The average fraction of the kinetic energy of the beam that is
radiated in these photons is 
 
\begin{equation}
\xi _\gamma \equiv \frac{\left\langle \Phi _\gamma \right\rangle }{
n_0m_pc^3\Gamma _0}
\simeq 10^{-3}\left( {\frac{\Gamma _0}{10^2}}\right)
^2\left( {\frac{B_0}{10^3\,\rm{G}}}\right)\,,  
\label{ksinum}
\end{equation}
 
\noindent where $\left\langle\Phi_\gamma \right\rangle$
is the average synchrotron luminosity of
high-energy electrons per unit area of the barrier front. 

In the burster frame, the characteristic energy, $\langle \tilde{\varepsilon}_
\gamma \rangle$, of synchrotron photons generated in the vicinity of the 
wind front increases due to the Doppler effect. Taking this
into account and using equation (16), we have 

\begin{equation}
\langle \tilde{\varepsilon}_\gamma \rangle \simeq
10\left( {\frac{\Gamma _0 }{10^2}}\right) ^2\left(
{\frac{B_{\rm{dec}}}{10^5\,\rm{G}}}\right) \;\;\rm{ MeV}\,.
\label{epsLab}
\end{equation}

\noindent
The fraction of the wind energy which is transferred to radiation at the 
wind front does not depend on the frame where it is estimated,
and in the burster frame it is equal to

\begin{equation}
\tilde{\xi}_\gamma = \xi_\gamma\simeq  10^{-3}\left( \frac{\Gamma
_0}{10^2} \right) \left( \frac{B_{\rm{dec}}}{10^5\, \rm{G}}\right) \,,
\label{ksiLab} 
\end{equation}
 
\noindent 
For typical parameters of relativistic magnetized winds which 
are relevant to cosmological $\gamma$-ray bursters,
$\Gamma_0\simeq 10^2-10^3$ and $B_{\rm{dec}}\simeq 10^5$ G,
from equations (18) and (19)
we have $\langle \tilde{\varepsilon}_\gamma \rangle\simeq 10 - 10^3$
MeV and $\tilde{\xi}\simeq 10^{-2}-10^{-3}$. Hence, the synchrotron
radiation that is generated at wind front may be responsible  for high-energy
$\gamma$-rays that are observed in the spectra of some GRBs
(e.g., Hurley et al. 1994). 

The main part of the X-ray and $\gamma$-ray emission of
detected GRBs is in the BATSE range, from a few $\times 10$ keV to a few
MeV (e.g., Fishman \& Meegan 1995). Synchrotron radiation from the wind front is either 
too hard or too weak to explain this emission irrespective of $B_{\rm dec}$. 
Indeed, equations (18) and (19) yield
 
\begin{equation}
\langle \tilde{\varepsilon}_\gamma \rangle \simeq 10^2\left(
{\frac{\Gamma _0}{10^2}}\right) \left( {\frac{\tilde{\xi}_\gamma
}{10^{-2}}} \right)\;\;\;\rm{MeV}.  
\label{epsLabksiLab}
\end{equation}

\noindent In our model, the energy of rotation powered winds
which are responsible for cosmological GRBs cannot be significantly 
more than $10^{53}$ ergs. To explain the energy output of $\sim 10^{51}-
10^{53}$ ergs per GRB, the  efficiency of transformation
of the wind energy into the energy of non-thermal radiation must be
more than $ 1\%$, $\tilde{\xi}_\gamma > 10^{-2}$, and maybe, for some
GRBs it is as high as 100\%, $\tilde{\xi}_\gamma \simeq 1$. Taking into 
account that for cosmological GRBs the value of $\Gamma _0$ is 
$> 10^2 $ (e.g., Fenimore et al. 1993, Baring \& Harding 1997), 
for $\tilde{\xi}_\gamma > 10^{-2}$ from equation 
(20) it follows that the mean energy of synchrotron
photons generated at the wind front is very high, $\langle
\tilde{\varepsilon}_\gamma\rangle > 100$ MeV. This is because
the bulk of these photons is generated in the thin vicinity 
of the wind front where there are both a strong magnetic field, 
$B\simeq B_0$, and high-energy electrons. The mean time that
high-energy electrons spend in this vicinity and generate 
synchrotron radiation is very short, $\sim$ a few $\times T_p\sim
10^{-6}(B_{\rm dec}/10^5\,\,{\rm G})^{-1}(\Gamma_0/10^2)$ s.
To get a high efficiency of synchrotron radiation at the wind front, 
$\tilde{\xi}_\gamma > 10^{-2}$, it is necessary to assume that 
the magnetic field $B_{\rm dec}$ is about its maximum value,
$B_{\rm dec}\sim 10^6$ G. From equation (18), we can see that
in this case the mean energy of synchrotron photons is 
of the order of or higher than $10^2$ MeV.
 
For a typical value of $B_{\rm dec}$, that is $\sim 10^5$ G,
and $\Gamma_0\sim 10^2-10^3$, from equation (19)
we have $\tilde \xi_\gamma\sim 10^{-3}-10^{-2}\ll 1$.
Hence, electrons of the ambient medium are accelerated at 
the wind front and injected into the region ahead of the front
practically without energy losses. In our model,
these high-energy electrons are the best candidates to be
responsible for the X-ray and $\gamma$-ray emission of GRBs in the 
BATSE range.

\subsection{LAEMWs generated at the wind front}

\begin{figure}
\centerline{\epsfig{file=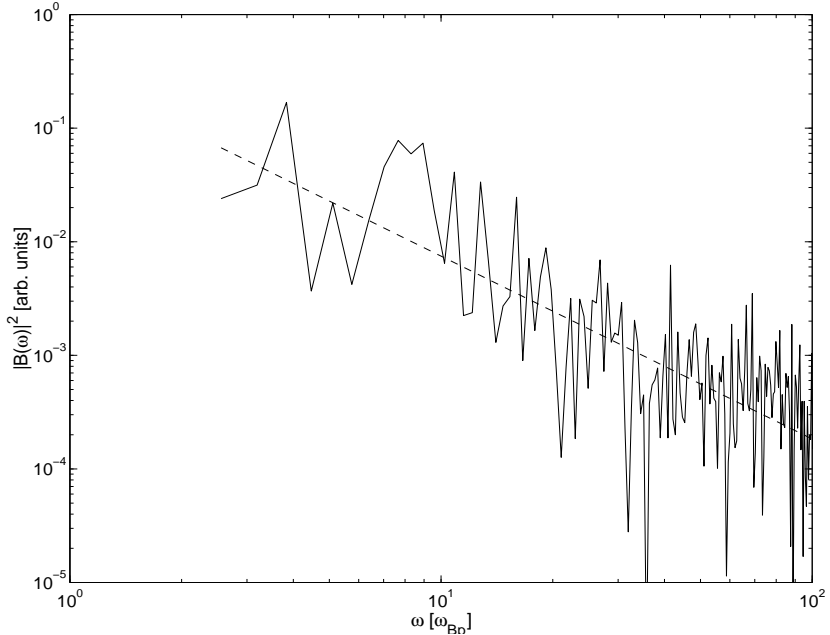,angle=90,width=11cm,height=8.5cm}}
\caption{Power spectrum of LAEMWs (thin line)
generated at the front of the magnetic barrier in 
a simulation with $B_0=300$ G, $\Gamma_0=300$ and $\alpha =2/3$.
The spectrum is fitted by a power law (dotted line).
} \label{fig4_usov}
\end{figure}

At $\alpha\sim 1$, LAEMWs are 
generated at the wind front and propagate in both 
directions from the front (see  Fig. 2). 
These waves are non-monochromatic. Figure~\ref{fig4_usov} shows a 
typical spectrum of LAEMWs in the wind frame. 
This spectrum has a maximum at the frequency $\omega_{\rm max}$ 
which is a few times higher than the proton
gyrofrequency $\omega_{Bp}=eB_0 /m_pc\Gamma_0$
in the field of $B_0$. 

At high frequencies, $\omega > \omega_{\rm max}$,
the spectrum of LAEMWs may be fitted by a power law:

\begin{equation}
|B(\omega )|^2\propto 
\omega ^{-\beta}\,,
\label{Bomega}
\end{equation}

\noindent
where $\beta\simeq 1.6$. 

The mean field of LAEMWs depends on $\alpha$, 
and in the wind frame, for $\alpha_{\rm cr} < \alpha  < 1$  
this field is

\begin{equation}
\langle B_w\rangle =(\langle B_z\rangle^2 + 
\langle E_y\rangle^2)^{1/2}\simeq 0.1 B_0 \simeq 0.1 B_{\rm dec}
/\Gamma_0
\label{avBw}
\end{equation}

\noindent
within a factor of 2 or so. 

\subsection{High-energy electrons accelerated at the wind front}

\begin{figure}
\centerline{\epsfig{file=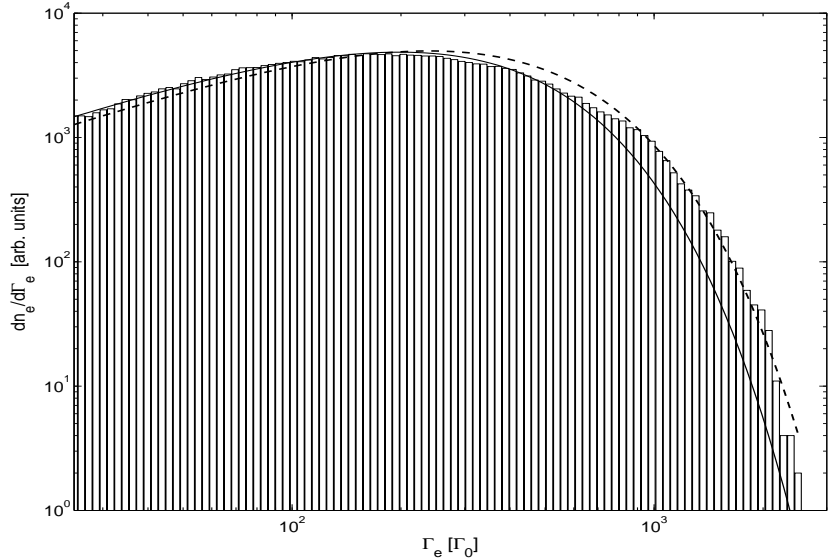,angle=90,width=11cm,height=7.5cm}}
\caption{Energy spectrum of highly accelerated electrons
in the region ahead of the wind front in the frame of the wind
for a simulation with $B_0=300$ G, $\Gamma_0=300$ and $\alpha =2/3$.
The electron spectrum is fitted by a two-dimensional relativistic
Maxwellian with a relativistic temperature $T=m_ec^2\Gamma_{_T}/k$,
where $\Gamma_{_T}$ is equal to either $240\Gamma_0$
(dotted line) or $200\Gamma_0$ (solid line).
} \label{fig5_usov}
\end{figure}

At $r\sim r_{\rm dec}$, where $\alpha$ is $\sim 1$, about $20$ \%
of the energy of a relativistic strongly magnetized wind
is transferred  to electrons of the ambient medium which are reflected 
from the wind front and accelerated to extremely high energies
(Smolsky \& Usov 1996, 1999, Usov \& Smolsky 1998). 
In the wind frame, the spectrum 
of high-energy electrons in the region ahead of the wind front 
may be fitted by a two-dimensional relativistic Maxwellian 
(Smolsky \& Usov 1999)

\begin{equation}
{dn_e \over d\Gamma_e}\propto \Gamma_e\exp \left({-
{m_ec^2\Gamma_e\over kT}}\right)
\label{dne}
\end{equation}

\noindent
with a relativistic temperature $T=m_ec^2\Gamma_{_T} /k$, where 
$\Gamma_{_T}\simeq 240 \Gamma_0$ (Fig.~\ref{fig5_usov} ). The fact that the  
energy distribution of accelerated electrons at $\alpha\sim 1$  
is close to a relativistic Maxwellian is quite natural because at $\alpha 
> \alpha_{\rm cr}$ the trajectories of particles in the
front vicinity are fully chaotic. 
The thermal distribution of high energy electrons does not
come as a result of interparticle collisions, since the ambient
medium is collisionless and no artificial viscosity is included in
the simulation code (Smolsky \& Usov 1996). Thermalization of the
electron distribution is purely a result of collisionless interactions
between particles and electromagnetic oscillations generated at the wind front.

At $\Gamma_e\leq 700
\Gamma_0$, the fit of the electron spectrum with a two-dimensional 
relativistic Maxwellian (23) with $\Gamma_{_T}
\simeq 200\Gamma_0$ is  rather accurate (Fig. 5). 
In this case, a small excess of electrons
with Lorentz factors $\Gamma_e > 700 \Gamma_0$ may be interpreted as 
a high-energy tail. Such a tail may result, for example,  
from multiple acceleration of high-energy electrons at the wind front.
The angular distribution of high-energy electrons in the 
region far ahead of the wind front is anisotropic. The mean angle
between the velocity of outflowing electrons and the normal to 
the wind front is $\langle \psi \rangle\simeq 1/3$ radian. 
 
\subsection{Synchro-Compton radiation of high-energy electrons
from the region ahead of the wind front}

High-energy electrons with a nearly Maxwellian spectrum are injected 
into the region ahead of the wind front and radiate in the fields of 
LAEMWs via synchro-Compton mechanism. For 
$\langle \psi \rangle\simeq 1/3$,
synchro-Compton radiation of high-energy electrons closely
resembles synchrotron radiation (e.g. Blumenthal \& Tucker 1974,
Smolsky \& Usov 1999). Therefore, to model the
spectrum of synchro-Compton radiation, we replace the fields 
of LAEMWs $B_z(t,x)$ and $E_y(t,x)$ by a constant magnetic field
which is equal to the mean field of LAEMWs $\langle B_w\rangle$.
The energy losses of electrons in such a magnetic field are 
governed by (Landau \& Lifshitz 1971)

\begin{equation}
  \frac{d\Gamma_e}{dt}=-\chi (\Gamma_e^2 -1)\,,
  \label{SynchLoss}
\end{equation}

\noindent where $\chi = 2e^4\langle B_w\rangle^2/3m_e^3c^5$.

In our approximation, the evolution of the spectrum
of high-energy electrons in the region ahead of the 
wind front may be found from the following
equation (e.g., Pacholczyk 1969)

\begin{equation}
{\partial f(\Gamma_e, t)\over \partial t}=
{\partial \over \partial\Gamma_e}[\chi \Gamma_e^2
 f(\Gamma_e, t)] + \dot N _ef_{\rm bar}(\Gamma_e)\,,
\label{par}
\end{equation}

\noindent where $f(\Gamma_e, t)$ is the distribution function
of high-energy electrons in the region ahead of the 
wind front per unit area of the front at the moment $t$, 
$\dot N_e\simeq n_0c$ is the rate of production of high-energy 
electrons per unit area of the front, and 
$f_{\rm bar}=(\Gamma_e/\Gamma_{_T}^2)\exp (-\Gamma_e/
\Gamma_{_T})$ is the average spectrum of high-energy electrons
which are injected into the region ahead of 
the front. The function  $f(\Gamma_e, t)$ is 
normalized to the total number of high-energy electrons
per unit area of the front $N_e$, while the function $f_{\rm bar}$
is normalized to unity:

\begin{equation}
\int_1^{\infty}f(\Gamma_e, t)\,d\Gamma_e=N_e\,\,\,\,\,\,\,
{\rm and }\,\,\,\,\,\,\,\int_1^{\infty}f_{\rm bar}\,d\Gamma_e=1\,.
\label{norm}
\end{equation}

\noindent
For simplicity, we disregard the angular anisotropy of the electron 
distribution.

Under the mentioned assumptions, in the frame of the wind front
the differential proper intensity of synchro-Compton radiation 
from the region ahead of the wind front
is (e.g., Pacholczyk 1969, Rybicki \& Lightman 1979)

\begin{equation}
I_\nu (t)= \int_1^{\infty}f(\Gamma_e, t)i_\nu d\Gamma_e\,,
\label{Inu}
\end{equation}

\noindent where 

\begin{equation}
i_\nu =\frac{\sqrt{3}e^3\langle B_w\rangle}{m_ec^2}
\frac \nu {\nu _c}\int_{\nu /\nu
_c}^\infty K_{5/3}(\eta )d\eta \,,  
\label{specti}
\end{equation}
 
\noindent 
is the spectrum of synchrotron radiation generated by
a single relativistic electron in a uniform magnetic field $\langle B_w\rangle$, 
$K_{5/3}$ is the modified Bessel functions of 5/3 order and
 
\begin{equation}
\nu _c={\frac{3e\langle B_w\rangle\Gamma^2_e}{4\pi m_ec}}
\label{nuTyp}
\end{equation}

\noindent
is the typical frequency of synchrotron radiation.

The observed spectral flux $F_\nu (t)$ (in units erg s$^{-1}$
cm$^{-2}$ erg$^{-1}$) can be obtained from the proper
intensity $I_\nu (t)$ dividing by the square of the burster
distance $d$, and by taking into account both the effects of
relativistic beaming and cosmological effects
(e.g., Tavani 1996a, Dermer 1998):

\begin{equation}
F_\nu (t)={D^3(1+z)\over 4\pi d^2}I_{\nu'}(t)\,,
\label{Fnu}
\end{equation}

\noindent where $D$ is the relativistic Doppler factor, 
$D\simeq 2\Gamma_0$, and $z$ is the cosmological redshift.
The observed frequency $\nu$ depends on the emitted frequency
$\nu'$ in the wind frame as $\nu=[D/(1+z)]\nu'\simeq [2\Gamma_0/
(1+z)]\nu'$. 

\begin{figure}
\centerline{\epsfig{file=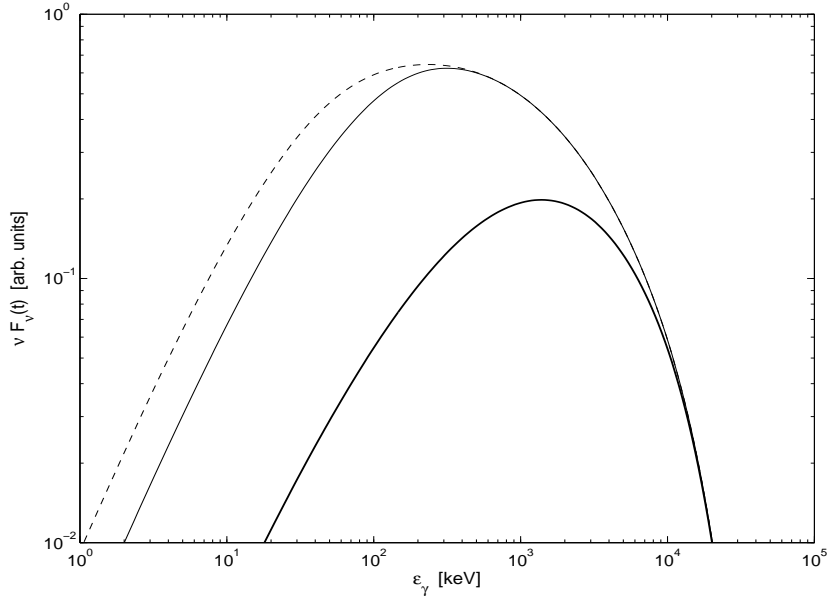,angle=90,width=11cm,height=8cm}}
\caption{Calculated spectral power of synchro-Compton radiation 
from the region ahead of the wind front as a function of photon energy
for $\Gamma_0 = 150$, $B_0= 300$ G, $\langle B_w\rangle =
0.1 B_0$, $\Gamma_T = 200\Gamma_0 = 3\times 10^4$ and $z=1$.
The spectrum of radiation is given for the moments when the fraction of
the energy of high-energy electrons injected into the region ahead of 
the wind front which is radiated is very small,
$\ll 1$\% (thick solid line), 36\% (thin solid line) and
58\% (dotted line).
} \label{fig6_usov}
\end{figure}

Equations (24) - (30) were integrated numerically.
For typical parameters of cosmological $\gamma$-ray
bursters, $\Gamma_0 = 150$, $B_0= 300$ G, $\langle B_w\rangle =
0.1 B_0$, $\Gamma_T = 200\Gamma_0 = 3\times 10^4$ and $z=1$, 
Figure~\ref{fig6_usov} shows the observed spectrum of 
synchro-Compton radiation
from the region ahead of the wind front.
For this radiation, the characteristic 
energy of photons is in the BATSE range  (Band et al. 1993,  
Schaefer 1994, 1998,  Fishman \& Meegan 1995,  Preece et al. 1996).
The spectrum  of synchro-Compton radiation displays a continuous
hard to soft evolution, in agreement with observational data on GRBs 
(Bhat et al. 1994). 

To fit the observed spectra of GRBs,
for accelerated electrons it was usually taken either a 
three-dimensional relativistic Maxwellian distribution (Katz 1994a,b)
or a sum of a three-dimensional relativistic Maxwellian distribution 
and a suprathermal power-law tail (Tavani 1996a,b).
The thermal character of the electron distribution (or its part) is consistent 
with our results (see \S~5.3). 
In our simulations, we observe a two-dimensional relativistic 
Maxwellian distribution of accelerated electrons.  However, the difference
between 2D and 3D Maxwellian distributions does not
change the calculated spectrum of synchrotron radiation
significantly (Jones \& Hardee 1979, Rybicki \& Lightman 1979).

In many papers  (see, for a review Piran 1998),
the interaction of a relativistic wind with an ambient medium and
non-thermal radiation generated at $r\sim r_{\rm dec}$
were considered in the frame of
the conventional model which was based on the following assumptions.

\noindent
(1) Two collisionless shocks form: an external shock that
propagates from the wind front into the ambient medium,
and an internal shock that propagates from the wind front
into the inner wind, with a contact discontinuity at the wind front
between the shocked material.

\noindent
(2) Electrons are accelerated at the shocks to 
very high energies. 

\noindent
(3) The shocked matter acquires embedded magnetic fields.
The energy density of these fields is about the energy 
density of high-energy electrons accelerated at the shocks.

\noindent
(4) Highly accelerated electrons generate high-frequency (X-ray 
and $\gamma$-ray) radiation of GRBs via synchrotron mechanism.

\noindent
(5)  The efficiency of conversion of the wind energy into 
accelerated electrons and then to high-frequency
radiation of GRBs is about 10\%~$\sim 30$\%. 

The idea about formation of two collisionless shocks 
near the front of a relativistic wind  outflowing from a 
cosmological $\gamma$-ray burster is based mainly on both 
theoretical studies which have shown that 
collisionless shocks can form in a rarified plasma
(e.g., Tidman \& Krall 1971, Dawson 1983, Quest 1985)
and the fact that such shocks have been observed 
in the vicinities of a few comets and planets (e.g., Leroy et al. 1982,
Livesey, Kennel \& Russell 1982, Omidi \& Winske 1990).
Undoubtedly, collisionless shocks exist and can
accelerate electrons to ultrarelativistic energies in
many astrophysical objects such as supernova remnants and
jets of active galactic nuclei. However, for the wind parameters 
which are relevant to cosmological GRBs (see \S~2), formation of 
collisionless shocks in the vicinity of the wind front
is very questionable,
especially if $\Gamma_0$ is as high as $10^3$ or more
(e.g., Smolsky \& Usov 1996, 1999, Mitra 1996, Brainerd 1999). 
As to an internal shock, it cannot form in
a Poynting flux-dominated wind, in principle (e.g., Kennel, 
Fujimura \& Okamoto 1983, Kennel \& Coroniti 1984).

Our model of the wind $-$ ambient medium interaction qualitatively 
differs from the conventional model which is based on the assumption
that an external collisionless shock forms just ahead of the wind front.
Although it might seem that observational consequences of our model must 
differ from  observational consequences of the conventional model
significantly, this is not the case. Moreover,
we can see that all the listed assumptions of  the conventional model
except of the first one are confirmed by our simulations.
There is only a modification that LAEMWs are embedded in the region
ahead of the wind front instead of magnetic fields, and 
highly accelerated electrons generate high-frequency emission of GRBs 
via synchro-Compton radiation. However, in our case synchro-Compton 
radiation closely resembles synchrotron radiation. Therefore, 
if in the conventional model of GRB emission from the shocked region 
ahead of the wind front the mechanism of electron
acceleration by a relativistic collisionless shock is replaced by the
mechanism of electron acceleration at the wind front,
this model will remain otherwise practically unchanged.

At $r >r_{\rm dec}$, the outflowing wind slows down
due to its interaction with the ambient medium, and
when the Lorentz factor of the wind front is about several tens
or less, an external shock may form
just ahead of the front. We believe that the afterglows 
which are observed in $\sim 10^5$ s after some GRBs 
result from acceleration of electrons by such shocks
as it is generally accepted (M\'{e}sz\'{a}ros \& Rees 1997b,
Vietri 1997, Waxman 1997, Wijers, Rees \& M\'{e}sz\'{a}ros 1997).


\acknowledgments

This research was supported by MINERVA Foundation, Munich / Germany.


%
%

%

\end{document}